\documentclass[12pt]{article}
\input epsf
\usepackage{epsfig}

\def\mod{\mathrm{mod}}

\def\R{\mathbf{R}}

\def\C{\mathbf{C}}
\def\bC{\mathbf{\overline{C}}}
\begin{document}
\title{Quasi-exactly solvable quartic: real algebraic spectral locus}
\author{Alexandre Eremenko and Andrei Gabrielov\thanks{Both authors
are supported by NSF grant DMS-1067886.}}
\maketitle
\begin{abstract}
We describe the real quasi-exactly sol\-va\-ble spectral
lo\-cus of the PT-sym\-metric
quar\-tic
using the
Nevanlinna pa\-ra\-met\-ri\-za\-tion.
\end{abstract}


Following Bender and Boettcher \cite{BB}, we consider the eigenvalue problem
in the complex plane
\begin{equation}\label{phys}
w^{\prime\prime}+(\zeta^4+2b\zeta^2+2iJ\zeta+\lambda)w=0,
\quad w(te^{-\pi i/2\pm\pi i/3})\to 0,\quad
t\to+\infty,
\end{equation}
where $J$ is a positive integer.
This problem is quasi-exactly solvable \cite{BB}:
there exist $J$ elementary eigenfunctions
$w=p_n(\zeta)\exp(-i\zeta^3/3-ib\zeta)$,
where $p_n$ is a polynomial of degree $n=J-1$. 

When $b$ is real, the problem is PT-symmetric.
By the change of the independent variable $z=i\zeta$,
(\ref{phys}) is equivalent to
\begin{equation}\label{oper}
-y^{\prime\prime}+(z^4-2bz^2+2Jz)y=\lambda y,\quad y(te^{\pm\pi i/3})\to 0,\quad t\to+\infty.
\end{equation}
Polynomial $h$ in the exponent of an elementary eigenfunction $y(z)$ is 
$h(z)=z^3/3-bz$.
The {\em spectral locus} $Z_J$ is defined as
$$\{(b,\lambda)\in\C^2:\exists\; y\neq0\;\mbox{satisfying (\ref{oper})}\}.$$
The {\em real spectral locus} $Z_J(\R)$ is $Z_J\cap\R^2$. The {\em quasi-exactly
solvable spectral locus} $Z_J^{QES}$
is the set of all $(b,\lambda)\in Z_J$ for which
there exists an elementary solution $y$ of (\ref{oper}). This is a smooth
irreducible 
algebraic curve in $\C^2$, \cite{A,Bk}. In this paper we describe
$Z_J^{QES}(\R)=Z_J^{QES}\cap\R^2$.
We prove a result announced in \cite{EGp}:
\vspace{.1in}

\noindent
{\bf Theorem 1.} For $n\geq 0$,
{\em $Z_{n+1}^{QES}(\R)$ consists of $[n/2]+1$
disjoint analytic curves
$\Gamma_{n,m},\; 0\leq m\leq[n/2]$
(analytic embeddings of $\R$ to $\R^2$).

For $(b,\lambda)\in\Gamma_{n,m}$, the eigenfunction has $n$ zeros,
$n-2m$ of them real.

If $n$ is odd, then $b\to+\infty$ on both ends of each curve
$\Gamma_{n,m}$. If $n$ is even, then the
same holds for $m<n/2$, but on the ends of
$\Gamma_{n,n/2}$ we have $b\to\pm\infty$.

If $(b,\lambda)\in \Gamma_{n,m},\; (b,\mu)\in\Gamma_{n,m+1}$
and $b$ is sufficiently large, then $\mu>\lambda$.
}
\vspace{.1in}

This theorem establishes the main features of $Z_{n+1}^{QES}(\R)$ 
which can be seen in the computer-generated figure in \cite{BB}.
Similar results were proved in \cite{EG-M} for two other PT-symmetric
eigenvalue problems.

Our theorem parametrizes {\em all} polynomials $P$ of degree $4$
with the property that the differential equation $y^{\prime\prime}+Py=0$
has a solution with $n$ zeros, $n-2m$ of them real \cite{Shinl,EM,EG-M}.

Suppose that $(b,\lambda)\in Z_J^{QES}(\R)$.
Then the corresponding eigenfunction $y$ of (\ref{oper}) can be always
chosen real. Let $y_1$ be a real solution of the differential
equation in (\ref{oper}) normalized by $y_1(x)\to 0$ as $x\to+\infty,\;
x\in\R$. Then $y_1$ is linearly independent
of $y$. Consider the meromorphic function $f=y/y_1$. This function
has no critical points in $\C$, and the only singularities of $f^{-1}$ 
are six logarithmic branch points. A meromorphic function in $\C$ with
no critical points and whose inverse has finitely many logarithmic singularities
is called a {\em Nevanlinna function}.
All Nevanlinna functions $f$ arise from differential equations $y^{\prime\prime}+Py=0,$
where $P$ is a polynomial by the above construction: $f$ is a ratio of
two linearly independent solutions of the differential equation.

Consider the sectors
$$S_j=\{ te^{i\theta}:t>0,\;|\theta-\pi j/3|<\pi/6\},\quad j=0,\ldots,5.$$
The subscript $j$ in $S_j$ will be always understood as a residue modulo $6$.
Function $f$ has asymptotic values $\infty,0,c,0,\overline{c},0$
in the sectors $S_0,\ldots S_5$, where $c\in\bC$.
It is known that $f$ must have at least $3$ distinct asymptotic
values \cite{Nev}, so $c\neq 0,\infty$. Function $f$ is defined up
to multiplication by a non-zero real number,
so we can always assume that $c=e^{i\beta},\;
0\leq\beta\leq\pi$, where the points $0$ and $\pi$ can be identified.
The asymptotic value $c$ is called the  {\em Nevanlinna parameter}.
There is a simple
relation between $c$ and the Stokes multipliers \cite{Sibuya,Masoero}.

The sectors $S_j$ correspond to logarithmic singularities of the inverse function $f^{-1}$.
Thus
$f^{-1}$ has $6$
logarithmic singularities that lie over $4$ points if $c\neq{\overline{c}}$,
or over $3$ points if $c=\overline{c}$. 

The map $(b,\lambda)\mapsto \beta \;(\mod\,\pi),\; Z_J^{QES}(\R)\to\R$ is
analytic
and locally invertible \cite{Sibuya,Bk}, so
$\beta$ can serve as a local parameter on the real QES
spectral locus. To obtain a global parametrization one needs
suitable charts on $Z_J^{QES}(\R)$, where this map is injective.

To recover $f$, one has to know the asymptotic value $c$
and one more piece of information,
a certain cell decomposition of the plane described below.
Once $f$ is known, $b$ and $\lambda$ are found from the formula
\begin{equation}\label{schwarz}
\frac{f^{\prime\prime\prime}}{f^{\prime}}-\frac{3}{2}\left(
\frac{f^{\prime\prime}}{f^\prime}\right)^2=-2(z^4-2bz^2+2Jz-\lambda).
\end{equation}
Now we describe, following \cite{EG-CMPH},
the cell decompositions needed to recover $f$ from $c$.
Suppose first that $c\notin\R$. 
\bigskip
\begin{center}
\epsfxsize=5.5in%
\centerline{\epsffile{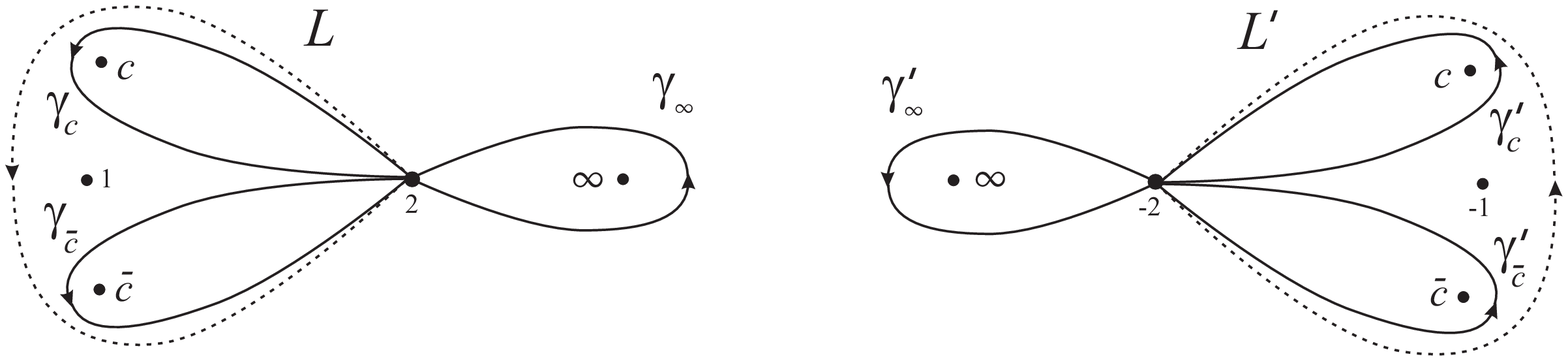}}
\nopagebreak\medskip
\noindent Fig. 1. Cell decompositions $\Phi$ and $\Phi'$ of the sphere
(solid lines).
\end{center}
\bigskip
Consider the cell decomposition $\Phi$
of the Riemann sphere $\bC$ shown by solid lines in the left part of Fig.~1.
It consists of one vertex at the point
$2$, three edges (loops $\gamma_c,\;\gamma_{\overline{c}}$ and $\gamma_\infty$
around non-zero asymptotic values)
and four faces (cells of dimension $2$). The faces are
labeled by the asymptotic values $0,c,\overline{c},\infty$. 
Label $0$ is not shown in the picture. The face labeled $0$ is
the unbounded region in the picture. 
(The point $1$ in the figure is neither
a label, nor a part of the cell decomposition. 
It will be needed, together with the dashed line $L$, for the limit at $\beta=0$.) 
As $$f:\C\setminus f^{-1}(\{0,\infty,c,\overline{c}\})\to\bC\setminus\{
0,\infty,c,\overline{c}\}$$
is a covering map, the cell decomposition $\Phi$ pulls back to
a cell decomposition $\Psi$ of the plane.
\bigskip
\begin{center}
\epsfxsize=4.0in%
\centerline{\epsffile{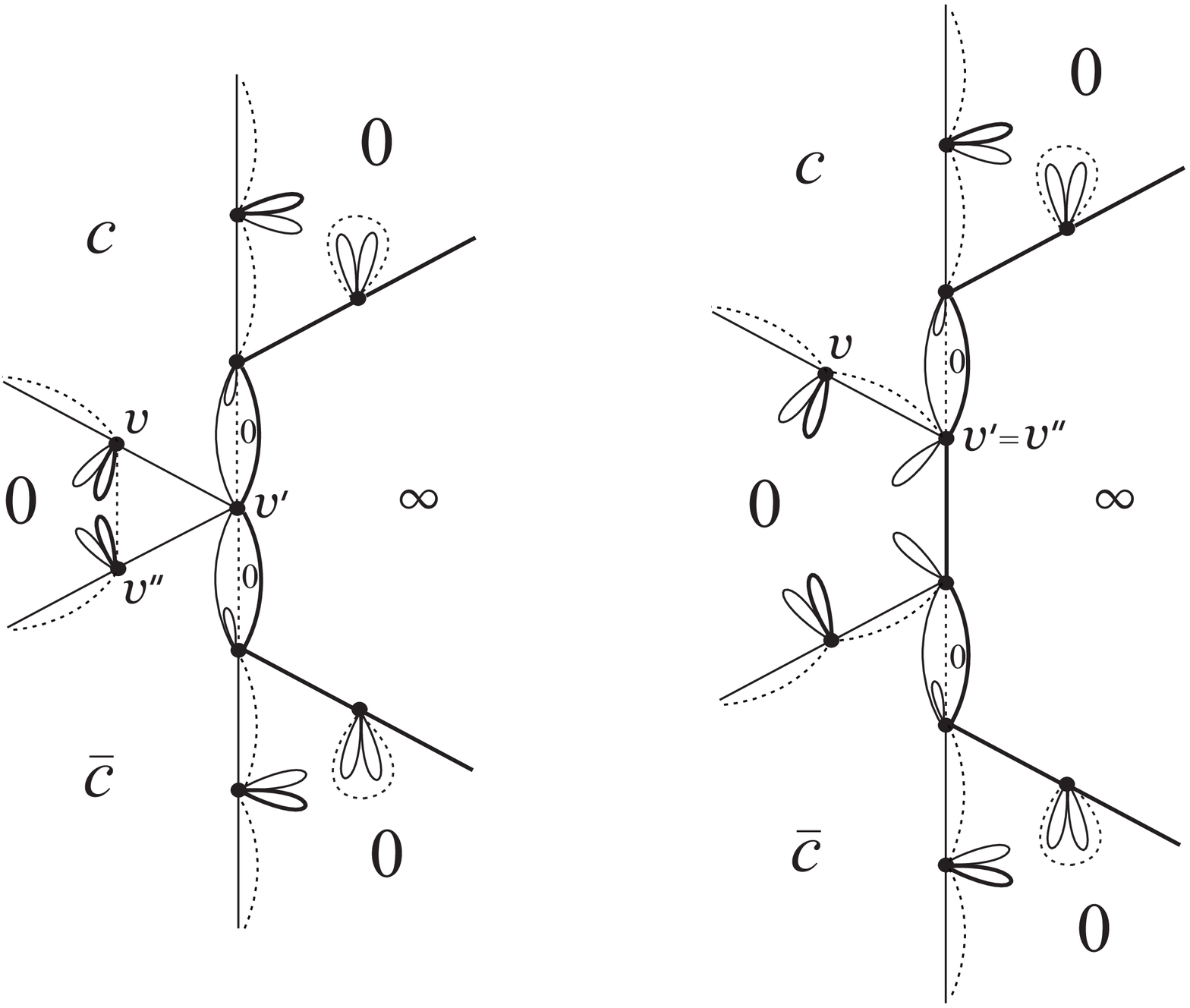}}
\nopagebreak\medskip
\noindent Fig. 2. Two examples of the cell decomposition $\Psi$ of the plane 
(solid lines). Both eigenfunctions have two zeros, none of them real.
\end{center}
\bigskip
Examples of $\Psi$ are shown with solid lines in Figs.~2 and 4 (left). 
The faces of $\Psi$ are labeled with the same labels as their images. 
Non-zero labels of bounded faces are omitted in the picture. 
The reader can restore them from the condition
that labels around a vertex must be in the same cyclic order as in Fig.~1 
(left, solid lines).
The labeled cell decomposition $\Psi$ defines $f$ up to a pre-composition
with an affine map of $\C$. Two cell decompositions define the same
$f$ if they can be obtained from each other by a 
homeomorphism of the plane preserving orientation and the labels.
Such cell decompositions are called equivalent.

By replacing multiple edges of the $1$-skeleton
of $\Psi$ with single edges and 
removing the loops, we obtain a simpler
cell decomposition $T$ whose $1$-skeleton is a tree, which we denote by
the same letter $T$. 
The cell decomposition $\Psi$ is uniquely recovered from its tree $T$ embedded
in the plane, 
\cite{EG-CMPH}.
The faces of $T$ are asymptotic to the sectors $S_j$ and the label of
each face is the asymptotic value in $S_j$. Two faces with
a common edge cannot have the same label. The cell decomposition $T$
is invariant under the reflection in the real axis,
with simultaneous
interchange of $c$ and $\overline{c}$. It is easy to classify all
possible embedded planar trees $T$ with labeled faces that satisfy these properties.
They depend on two integer parameters $k$ and $l\geq 0$.
These trees form two families, $X_{k,l},\; k\geq 0,l\geq 0$ 
and $X_{k,l},\;k<0,l\geq 0,$ shown in Fig.~3. Integers $|k|$ and $l$
are the numbers of edges between ramification vertices, as shown in Fig.~3. 

\bigskip
\begin{center}
\epsfxsize=4.5in%
\centerline{\epsffile{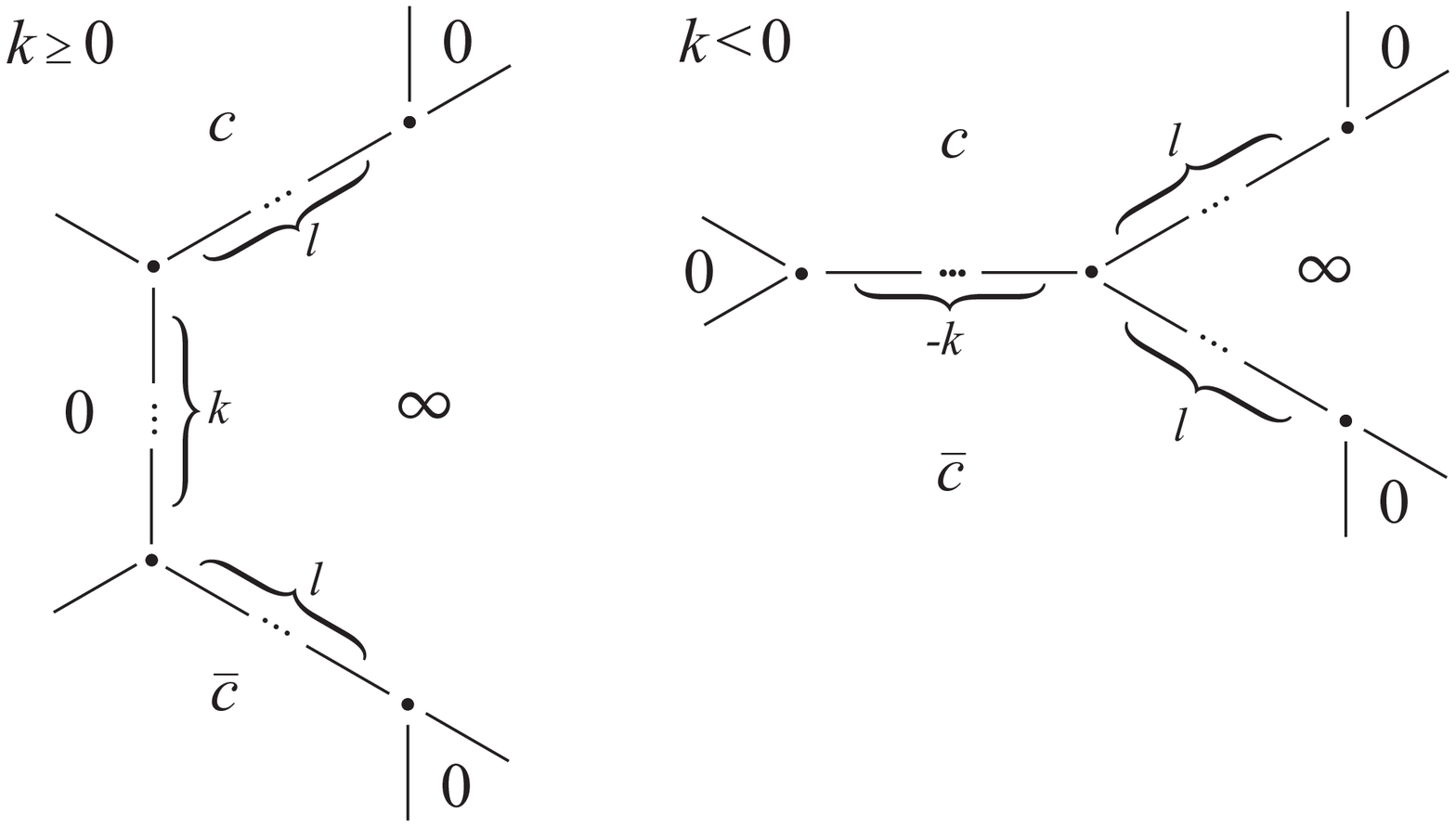}}
\nopagebreak\medskip
\noindent Fig. 3. Trees $X_{k,l}$. 
\end{center}
\bigskip
Cell decompositions in Fig.~2 (solid lines) correspond to the trees $X_{0,1}$
and $X_{1,1}$. Cell decomposition in the left part of Fig.~4 (solid lines)
corresponds to the tree $X_{-1,1}$ in the right part of Fig.~4. 

Parameters of the trees $X_{k,l}$ can be interpreted as follows:
$$k^-:=\min\{-k,0\}$$
is the number of real zeros of $f$,
and $2l$ is the number of non-real zeros. So the total number of zeros is
$n=2l+k^-.$

Functions $f$ corresponding to the trees $X_{k,l},\; k\geq 0$,
have $2l$ zeros, none of them real.
Zeros of the eigenfunction $y$ coincide with those of $f$.

For given $n$, the number of trees $X_{k,l}$ with $k<0,\; 2l-k=n$ is
\begin{equation}\label{eo}
(n+1)/2\; \mbox{when}\; n\; \mbox{is odd, and}\; n/2\;\mbox{when}\;
n\;\mbox{is even.}
\end{equation}
Every tree $X_{k,l}$ and every $\beta\in(0,\pi)$ defines a
meromorphic function $f$
satisfying (\ref{schwarz}) with $J=2l+k^-+1$ 
and some $(b,\lambda)$ depending on $\beta$, $k$ and $l$. 
This follows from a result of Nevanlinna \cite{Nev}, see also \cite{EG-CMPH}.
{}From this function $f$, the coordinates of a point $(b,\lambda)$ on
the real QES spectral locus are recovered from the Schwarz
equation (\ref{schwarz}).
Thus we have a map $F:(T,\beta)\mapsto (b,\lambda)$ which we call
the {\em Nevanlinna map.}
This map is of highly transcendental nature:
construction of $f$
from $T$ and $\beta$ involves the uniformization theorem.
We refer to \cite{EG-CMPH,EG-M,Nev} for details.

Each of the trees from our classification defines a chart of $Z_J^{QES}(\R)$.
To obtain the global parametrization of $Z_J^{QES}(\R)$, 
we only have to find out how these charts are pasted together.
We will see that the boundaries of our charts correspond to
the values $c=\pm1$.
When $c$ is real, we can use, instead of $\Phi$, the cell decomposition
$\Phi_1$ of the sphere with two loops,
$\gamma_\infty$ and the loop $L$ around $c$
shown with the dashed line in the left part of Fig.~1.
\vspace{.1in}

{\em Proof of Theorem 1.} We begin with the charts 
$X_{k,l},\; k<0$.
We show that in these charts the limits as $\beta\to 0,\,\pi$ do not
belong to the spectral locus.
This is proved by the arguments similar to those in
\cite[Thm. 4.1]{EG-M}.
\vspace{.1in}

\noindent
{\bf Lemma 1.} {\em For $k<0$ and $l\geq 0$, the limit of the
Nevanlinna map is
$$\lim_{\beta\to 0}F(X_{k,l},\beta)=\infty.\footnote
{Here $\infty$ refers to a point added to
the $(b,\lambda)$-plane in the one-point
compactification.}$$
}

{\em Proof.} When $\beta\to 0$, we have $c\to1,\;\overline{c}\to1$.
Suppose by contradiction that $F(X_{k,l},\beta)$ has a limit
$(b_0,\lambda_0)$. Then there is a limit function $f_0$, a solution of the
Schwarz equation (\ref{schwarz}) with these parameters $b_0$ and $\lambda_0$.
Meromorphic function $f_0$ has three asymptotic values, $0,1,\infty$,
and we are going to find the corresponding cell decomposition.
Let $\Phi_1$ be the cell decomposition of the Riemann sphere with one vertex
at the point $2$ and two loops, $\gamma_\infty$ and $L$ (see Fig.~1, left).
Let $\Psi_1=f^{-1}(\Phi_1)$.

It is easy to
construct $\Psi_1$ from the original cell decomposition $\Psi$.
First, removing preimages of $\gamma_c$ and $\gamma_{\overline{c}}$,
we obtain the cell decomposition $\Psi_\infty$, the preimage
of the loop around $\infty$ in $\Phi$.
It is shown with the bold solid lines in Fig.~4.

Next, for each vertex $v$ of $\Psi$ consider the path $L_v$
consisting of the edge of $f^{-1}(\gamma_c)$ starting at $v$ and 
ending at some vertex $v'$, followed by
the edge of $f^{-1}(\gamma_{\overline{c}})$ starting at $v'$ 
and ending at some vertex $v''$.
Then the edge of $f^{-1}(L)$ from $v$ to $v''$ is
homotopic to $L_v$ in the complement of $\Psi_\infty$.
The new edges are shown with dashed lines in Fig.~4.
The resulting cell decomposition is equivalent to $\Psi_1$.

Let $V$ be the set of the vertices of $\Psi$
contained in the boundary of the sector $S_3$.
It is connected to the rest of the vertices of $\Psi$
only at one vertex ($v'$ in Fig.~4, left) which is also
at the boundary of both sectors $S_2$ and $S_4$.
The dashed line replacing the edges of $\Psi$ that connect $v'$
to the two adjacent vertices of $V$
($v$ and $v''$ in Fig~4, left), goes from $v$ to $v''$.
All other dashed lines connect the vertices of
$V\setminus\{v'\}$ with the other vertices
from the same set. Hence $V\setminus\{v'\}$
is the set of vertices of a connected component
of the 1-skeleton of the cell decomposition $\Psi_1$.
This contradicts our assumption that $\Psi_1=f_0^{-1}(\Phi_1)$,
since the 1-skeleton of $f_0^{-1}(\Phi_1)$ must be connected.
This contradiction proves the lemma.
\bigskip
\begin{center}
\epsfxsize=4.5in%
\centerline{\epsffile{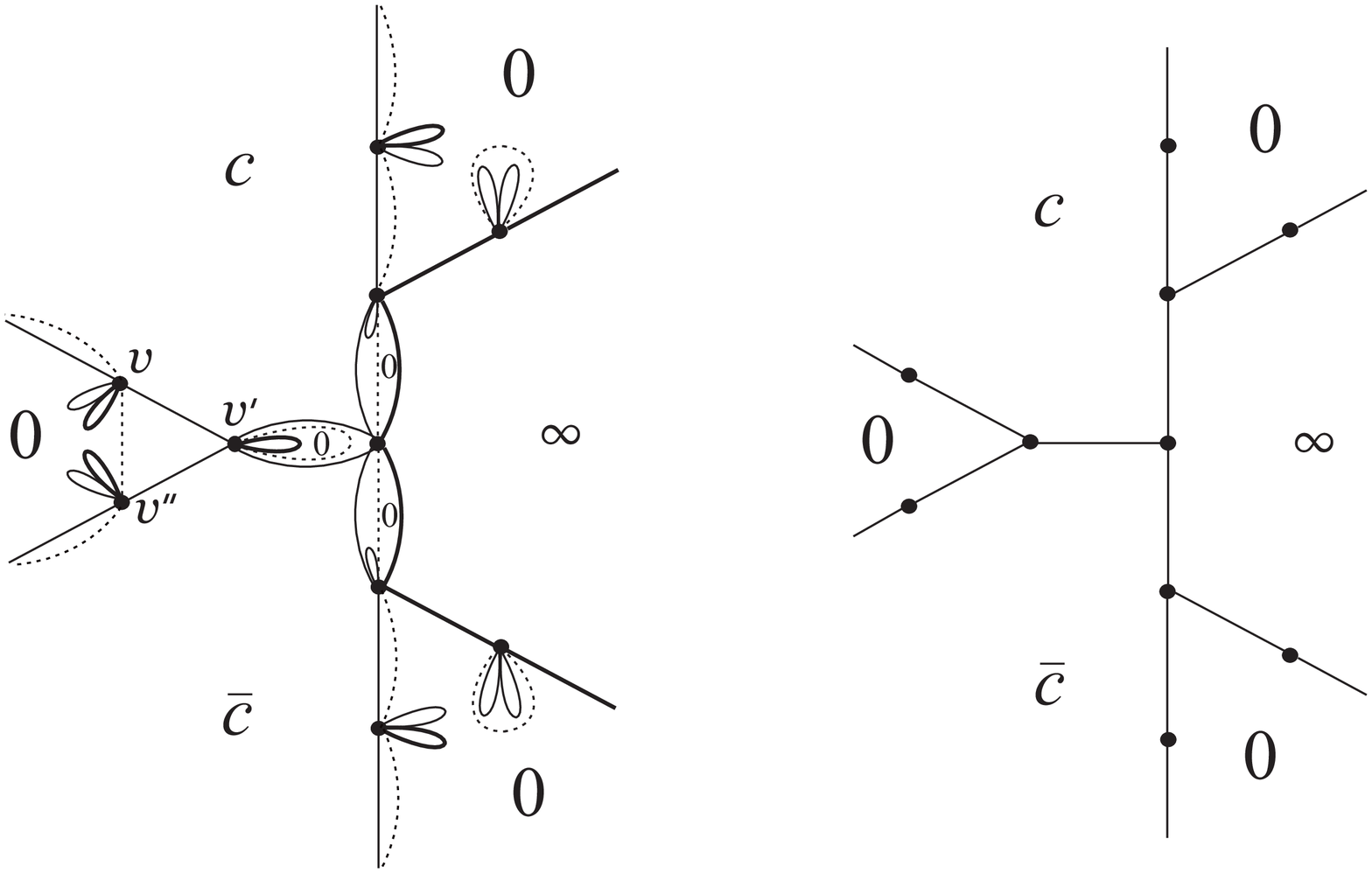}}
\nopagebreak\medskip
\noindent Fig.~4. A cell decomposition $\Psi$ (solid lines)
and the corresponding tree $X_{-1,1}$. Eigenfunction has one real
and two non-real zeros.
\end{center}
\bigskip
\vspace{.1in}

{\em Remark.} Consider all meromorphic functions with
no critical points and at most $6$ asymptotic values. These functions $f$
are defined by their asymptotic values and cell decompositions.
Assume that one vertex $v_0$ of $\Psi$ is placed at $z=0$ and
normalize so that $f'(0)=1$. The class of normalized functions
obtained in this way is {\em compact} \cite{Volk}.
Let $f_\nu\to f_0$ be a converging sequence.\footnote{Uniform convergence
on compact subsets in the plane, with respect to the spherical metric
in the target sphere.} The $1$-skeletons
of the corresponding cell decompositions
$\Psi(\nu)$ converge to the $1$-skeleton of the cell
decomposition $\Psi(0)$ as embedded graphs with a marked vertex.
If two asymptotic values collide in the limit, one has
to use the procedure described in the proof of Lemma 1:
replacing two loops by one loop.
The limiting cell decomposition obtained in Lemma 3 suggests
that the eigenvalue problem (\ref{oper}) tends to
a harmonic oscillator when $c\to 1$, the fact we'll later
prove by different arguments.
\vspace{.1in}

\noindent
{\bf Lemma 2.} {\em For $k<0$ and $l\geq 0$,
$$\lim_{\beta\to\pi}F(X_{k,l},\beta)=\infty.$$
}

{\em Proof.} When $\beta\to \pi$, we have $c\to-1,\;\overline{c}\to-1$.
Suppose by contradiction that $F(X_{k,l},\beta)$ has a limit
$(b_0,\lambda_0)$. Then there is a limit function $f_0$, a solution of the
Schwarz equation (\ref{schwarz}) with these parameters $b_0$ and $\lambda_0$.
Meromorphic function $f_0$ has three asymptotic values, $0,-1,\infty$,
and we are going to find the corresponding cell decomposition.

To do this, it is convenient to choose another cell decomposition $\Phi'$
of the Riemann sphere, shown in the right part of Fig.~1 (solid lines).
When $c\to-1$, $\Phi'$ collapses to $\Phi'_{-1}$ where the two loops 
$\gamma'_c$ and $\gamma'_{\overline{c}}$ 
are replaced with a single loop $L'$ around $-1$
(dashed line in Fig.~1, right).

We need the transition formula from 
$\Psi=f^{-1}(\Phi)$ to $\Psi'=f^{-1}(\Phi')$.
This formula is obtained by combining the two decompositions
(see Fig.~5) and expressing the loops of $\Phi'$ in terms
of the loops of $\Phi$.
\bigskip
\begin{center}
\epsfxsize=2.5in%
\centerline{\epsffile{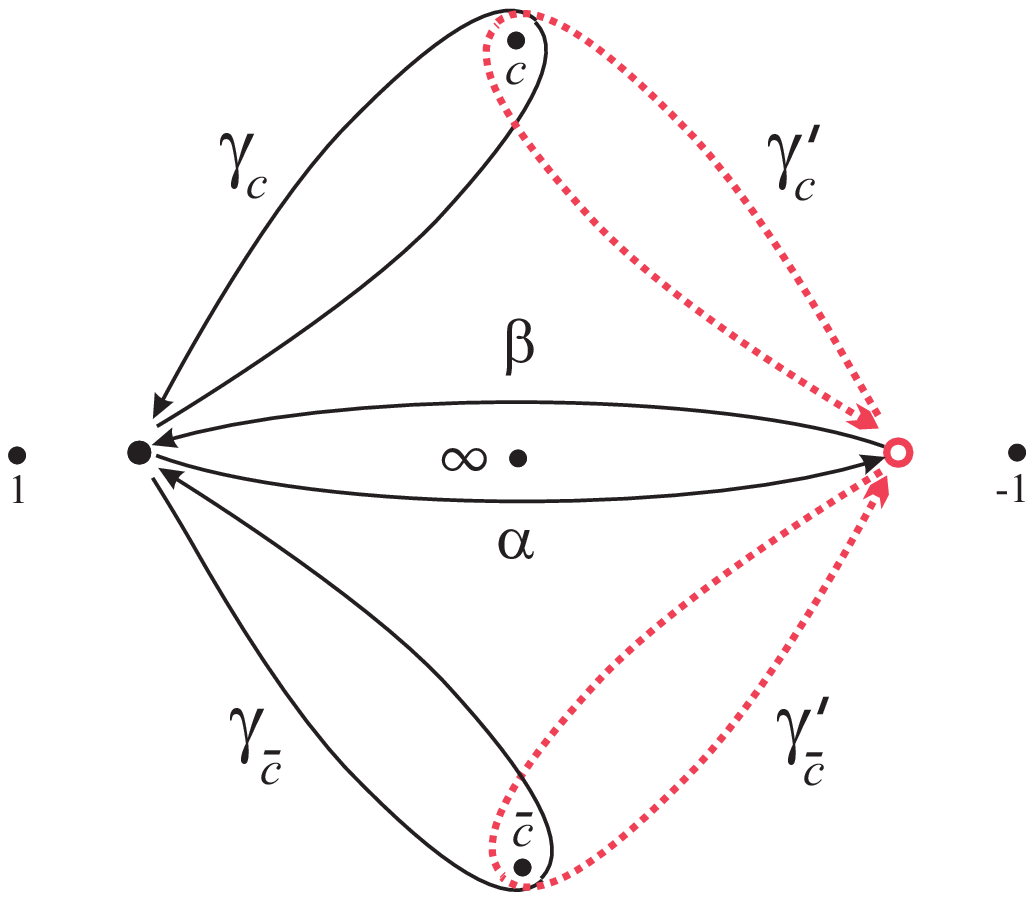}}
\nopagebreak\medskip
\noindent Fig.~5. Two cell decompositions of Fig.~1 combined.
\end{center}
\bigskip
The formulas, using notations in Fig.~5, are:
\begin{equation}\label{transition}
\gamma_\infty=\alpha\,\beta,\quad\gamma'_\infty=\beta\,\alpha,\quad
\gamma'_c=\beta\,\gamma_c\,\beta^{-1},
\quad\gamma'_{\overline{c}}=\alpha^{-1}\,\gamma_{\overline{c}}\,\alpha.
\end{equation}
Here the product should be read left to right.
Similar formulas were obtained in \cite{EG-M} in the proof of Theorem 4.1.
Application of these transition formulas
to the cell decomposition $\Psi$ of type $X_{-1,1}$ is illustrated
in Figs.~6,7.
\bigskip
\begin{center}
\epsfxsize=2.5in%
\centerline{\epsffile{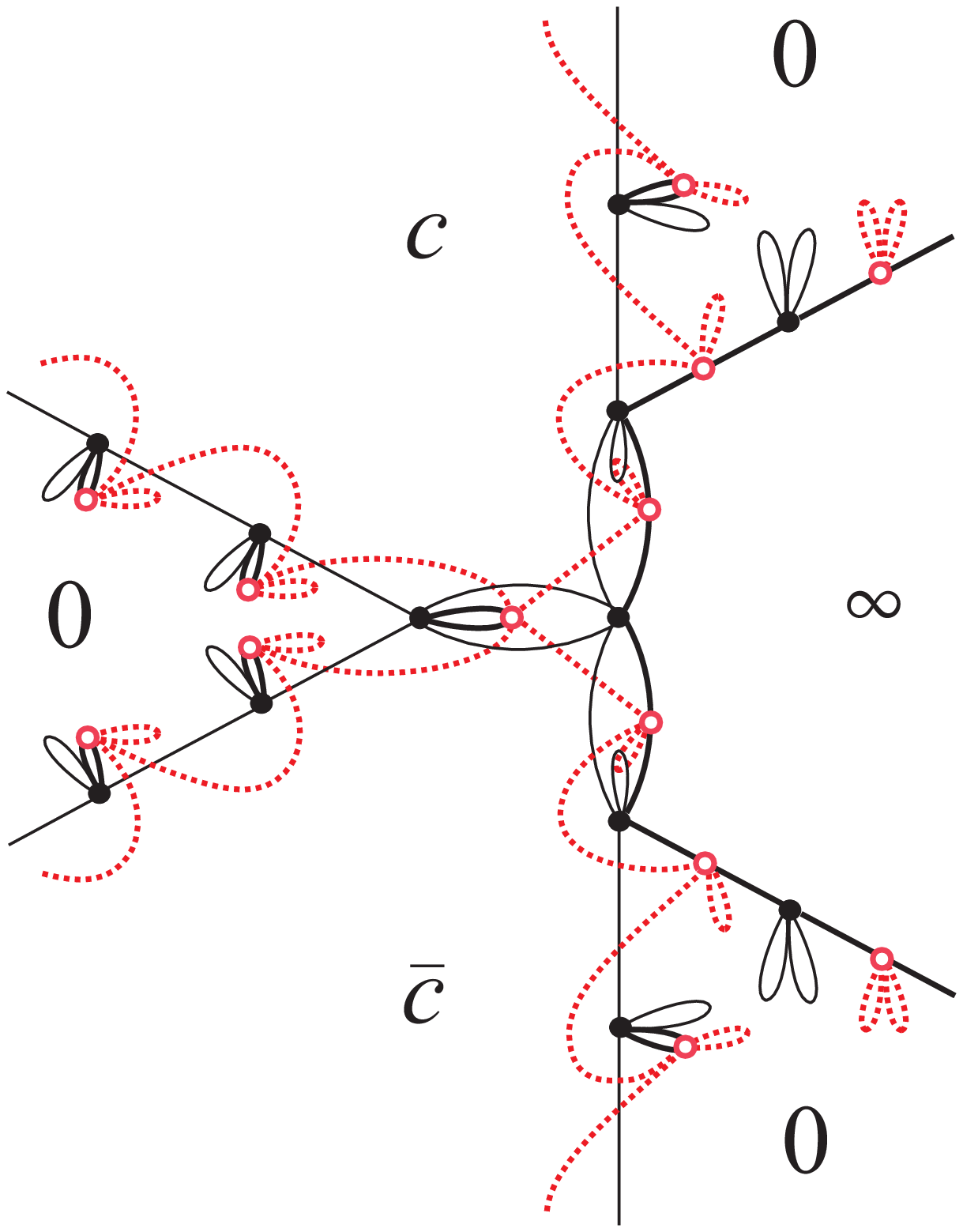}}
\nopagebreak\medskip
\noindent Fig. 6. Transition formulas (\ref{transition}) applied
to the cell decomposition in Fig.~4.
\bigskip
\epsfxsize=1.5in%
\centerline{\epsffile{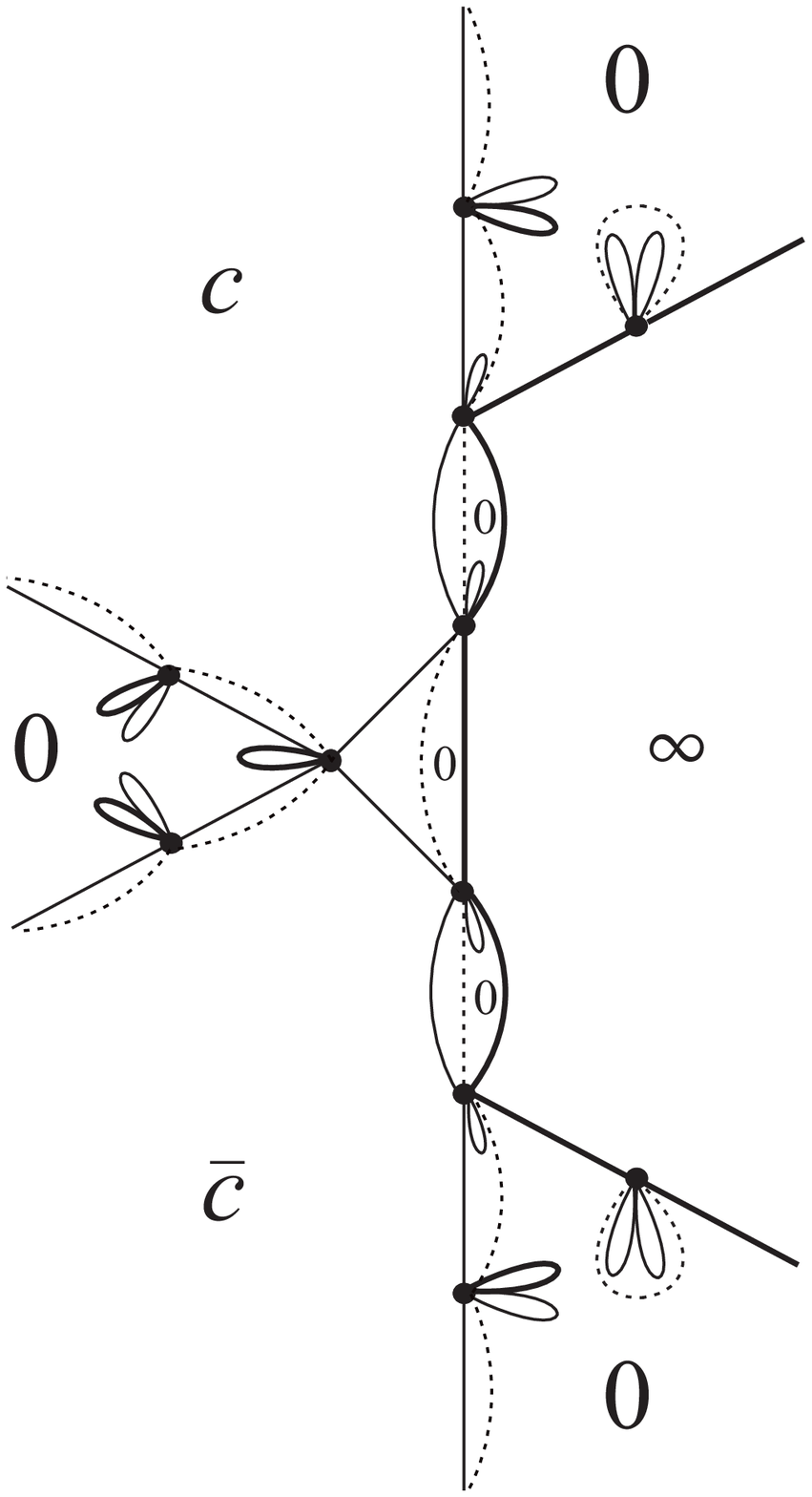}}
\nopagebreak\medskip
\noindent Fig. 7. Cell decomposition $\Psi'$. 
\end{center}
\bigskip
The same arguments as in Lemma 1 show that
the $1$-skeleton of the degeneration $\Psi'_{-1}$ of $\Psi'$ 
as $\beta\to\pi$ is not connected.
This proves the lemma.

Lemmas 1 and 2 show that for $k<0$, the charts $X_{k,m}$
with $2m-k=n$
cover connected components of $Z_{n+1}^{QES}(\R)$, each parametrized 
by $\beta\in(0,\pi)$.
We call these components $\Gamma_{n,m}$. 
These are simple disjoint analytically
embedded curves in $\R^2$. 

When $\beta\to 0,\pi$ we must have $b\to\pm\infty$.
We'll show below that $b\to+\infty$ on both ends of $\Gamma_{n,m}$
when $k>0$.

When $n$ is odd (that is $J$ is even), these curves $\Gamma_{n,m}$
constitute the whole spectral locus $Z_{n+1}^{QES}(\R)$.

Now consider the part of the spectral locus covered by the charts
$X_{k,l},\; k\geq 0$.
This part is present only when $n=2l$ is even.
\vspace{.1in}

\noindent
{\bf Lemma 3.} {\em For $k\geq 0$ and $l\geq 0$,
we have 
\begin{equation}\label{glu}
\lim_{\beta\to \pi}F(X_{k,l},\beta)=\lim_{\beta\to0}F(X_{k+1,l},\beta).
\end{equation}
and
\begin{equation}\label{gl1}
\lim_{\beta\to 0}F(X_{0,l},\beta)=\infty.
\end{equation}
}

{\em Proof of Lemma 3.} This is similar to the arguments in Lemmas 1 and 2.
Computation is illustrated in Figs.~2, 8, 9 and 10.
\bigskip
\begin{center}
\epsfxsize=4.0in%
\centerline{\epsffile{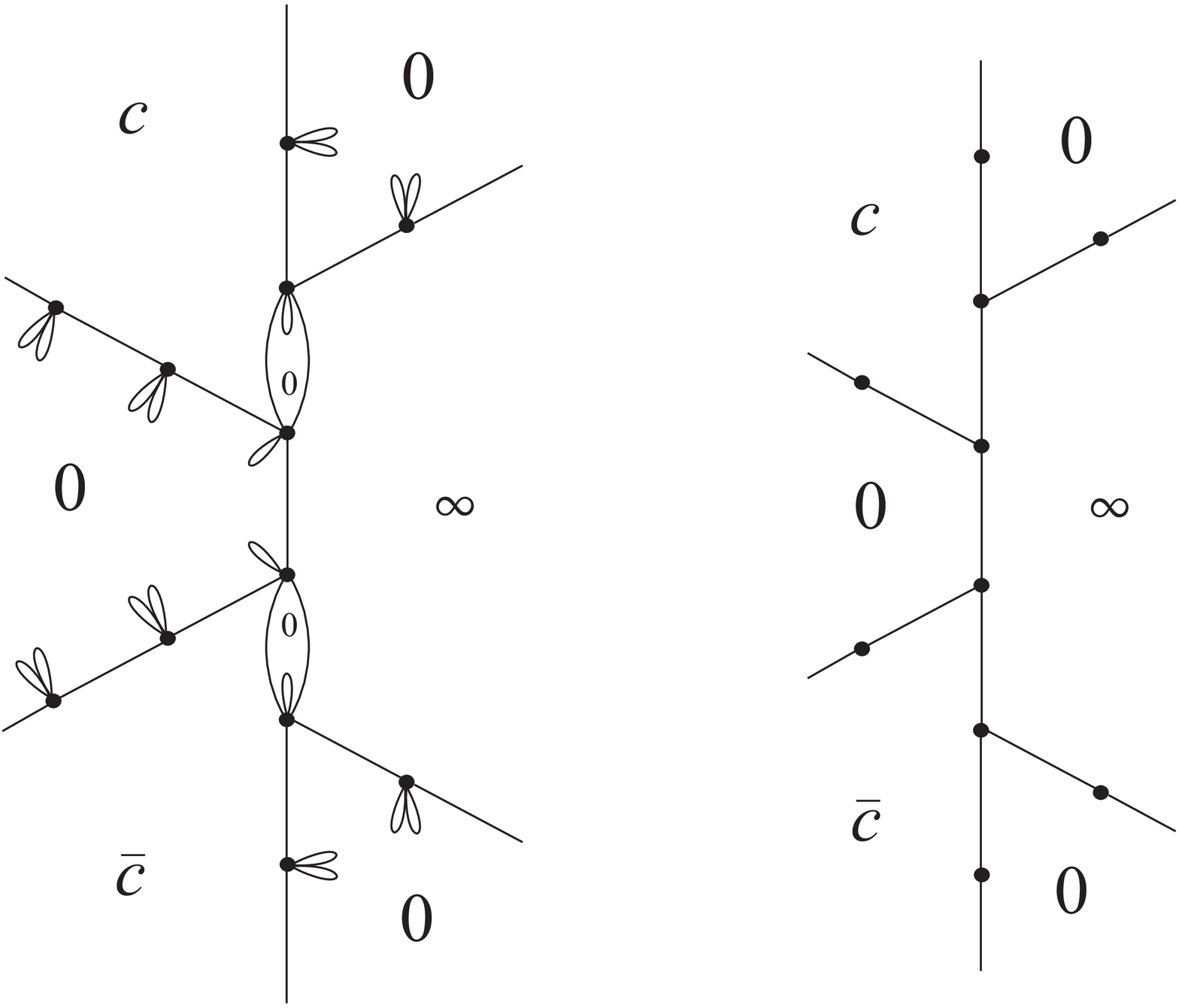}}
\nopagebreak\medskip
\noindent Fig.~8. Cell decomposition $\Psi$ corresponding to $X_{1,1}$.
\end{center}
\bigskip
In the left part of Fig.~8 we use $\Psi$ from Fig.~2, right.
It corresponds to the tree $X_{1,1}$ in Fig.~8, right.
In Fig.~9 the circles denote the vertices of $\Psi'$
(preimages of the vertex of $\Phi'$)
and the dotted lines correspond to the preimages of $\gamma'_c$ and 
$\gamma'_{\overline{c}}$ determined from (\ref{transition}).
The preimages of $\gamma_\infty$ and $\gamma'_\infty$ coincide.
They are shown with the bold solid line.
Removing the preimages of $\gamma_c$ and $\gamma_{\overline{c}}$
(thin solid lines in Fig.~9) and the vertices of $\Psi$,
we obtain the cell decomposition $\Psi'$ shown in Fig.~10 (left)
corresponding to the tree $X_{2,1}$ (right).
\bigskip
\begin{center}
\epsfxsize=2.0in%
\centerline{\epsffile{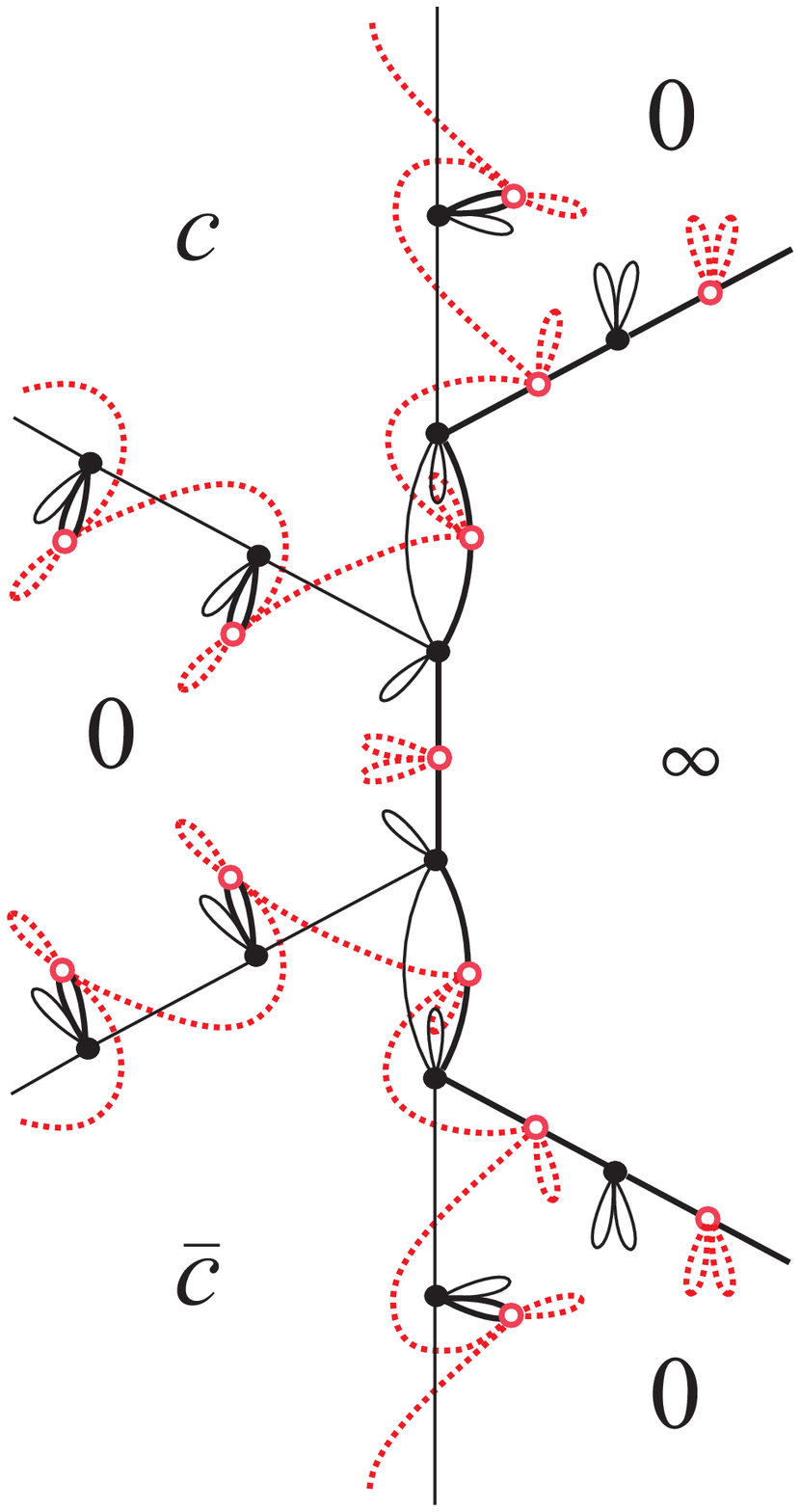}}
\nopagebreak\medskip
\noindent Fig. 9. Passing from $\Psi$ to $\Psi'$.
\end{center}
\bigskip
\begin{center}
\epsfxsize=4.0in%
\centerline{\epsffile{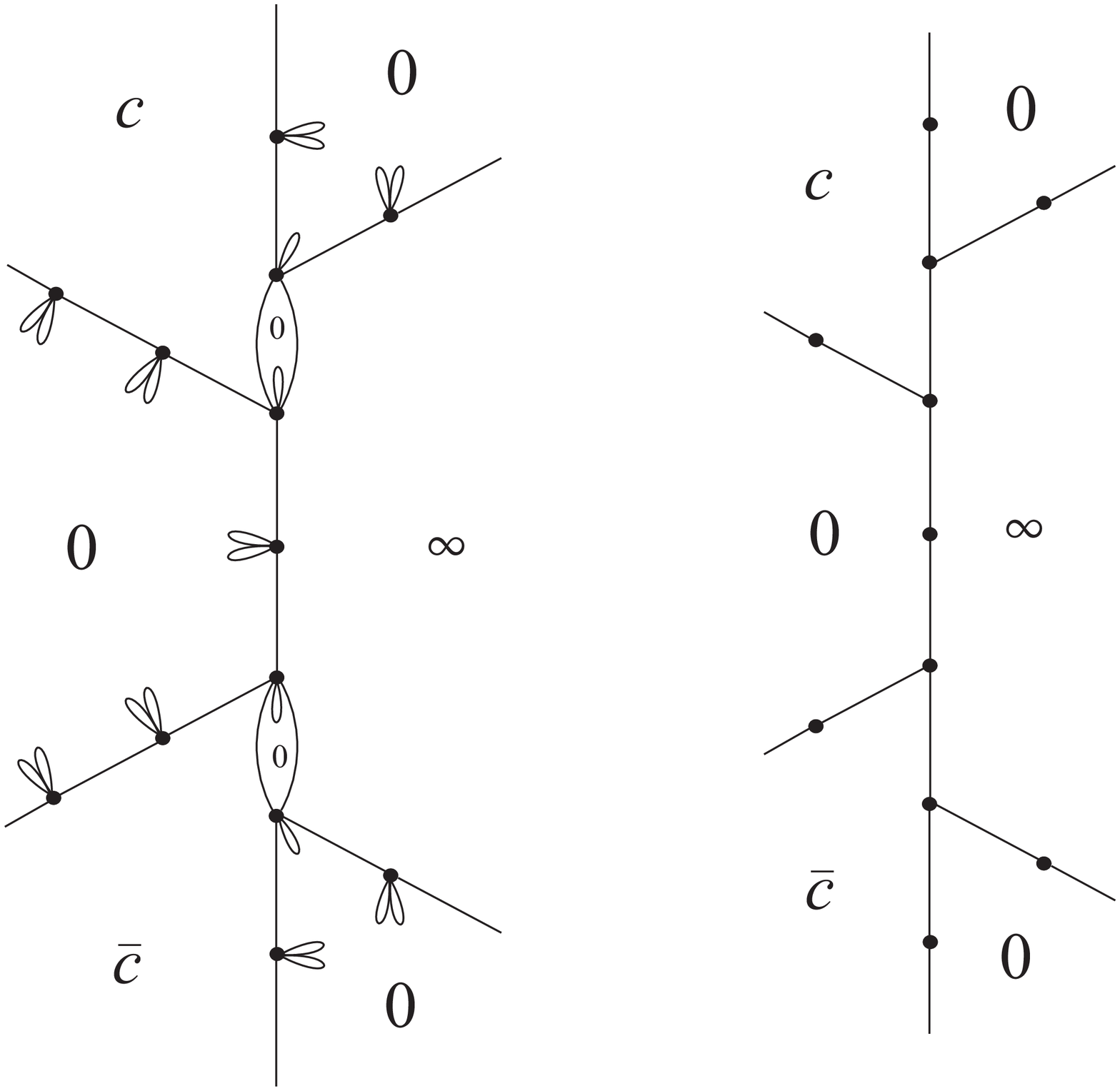}}
\nopagebreak\medskip
\noindent Fig. 10. Cell decomposition $\Psi'$ corresponding
to the tree $X_{2,1}$.
\end{center}
\bigskip
Thus, for $k\ge 0$, the cell decomposition $\Psi'_{-1}$ 
of the plane obtained from $X_{k,l}$
in the limit $\beta\to \pi$ as the preimage of $\Phi'_{-1}$
is equivalent to the cell decomposition $\Psi_1$ obtained from $X_{k+1,l}$
in the limit $\beta\to 0$ as the preimage of $\Phi_1$.
Since $\Phi'_{-1}=-\Phi_1$,
Nevanlinna theory implies that the corresponding functions $f$ and $f'$ 
satisfy $f'=-f$. Hence these two functions correspond to the same
point of $Z_J^{QES}(\R)$.

The proof of (\ref{gl1}) is similar to that of Lemma~1.
This completes the proof of the lemma.
\vspace{.1in}

Now we continue the proof of Theorem 1.

For even $n=2l$, charts $X_{k,l},\; k\geq 0$
parametrize segments of one curve in the real QES spectral locus,
and we call this curve $\Gamma_{n,n/2}$. We parametrize the curve
$\Gamma_{n,n/2}$ by the real line, so that the number $k$ decreases,
thus the right end
of $\Gamma_{n,n/2}$ is parametrized by the chart $X_{0,n/2}$.
So when the parameter $t\in\R$ on $\Gamma_{n,n/2}$ tends to $+\infty$, 
the asymptotic
value $c=\exp(i\beta)$ tends to $1$. On the other hand, when $t\to-\infty$
on $\Gamma_{n,n/2}$ the asymptotic value $c$ does not have a limit; 
it oscillates, passing each point of the unit circle infinitely many times.

The curves $\Gamma_{n,m}$ are disjoint. Indeed, different cell decompositions
give different functions $f$.
This proves the first two statements of Theorem 1.

Now we deal with the asymptotic behavior of our curves $\Gamma_{n,m}$.
We use the rescaling of (\ref{oper}) as in \cite{EGp}.
The QES spectral locus is defined by a polynomial
equation $Q_{n+1}(b,\lambda)=0$
which is of degree $n+1$ in $\lambda$. So on a ray $b>b_0$
there are $n+1$ branches $\lambda_j(b)$. 
In \cite[Eq. (25)]{EGp}, we found that all $\lambda_j$
have asymptotics $\lambda(b)\sim b^2+O(\sqrt{b}),\; b\to\infty$,
and as $b\to+\infty$, each QES eigenfunction $y_j$
tends to some eigenfunction $Y_\ell$
of the harmonic oscillator
\begin{equation}\label{harm}
-Y^{\prime\prime}+4z^2Y=\mu Y,\quad Y(it)\to 0,\quad t\to\pm\infty.
\end{equation}
The eigenvalues of this harmonic oscillator are
$\mu_\ell=2(2\ell+1),\quad \ell=0,1,\ldots.$

Only one of the eigenfunctions $y_j$ can tend to a given $Y_\ell$,
and the corresponding eigenvalue satisfies 
$$\lambda_j(b)=b^2+(\mu_\ell-2J+o(1))\sqrt{b},\quad b\to+\infty.$$
It follows that all $\lambda_j$ are real. The graph
of each $\lambda_j$ is a part of a curve $\Gamma_{n,m}$,
and each $\Gamma_{n,m}$ has only two ends.

Now we consider the degeneration of the $X_{0,l}$ chart with $l\geq 0$,
the chart which parametrizes the right end of $\Gamma_{n,n/2},\; n=2l.$
On the left end of $\Gamma_{n,n/2}$,
where $t\to-\infty$ in the parametrization described after Lemma 3, 
there are infinitely many points $\Gamma_{n,n/2}(t_k)$ which belong to
the real QES locus, and where the asymptotic value $c$ is real.
It was proved in \cite{EGp} that these are exactly those points where
$Z_J^{QES}(\R)$ crosses the non-quasi-exactly solvable part of $Z_J(\R),$
and these points correspond to $b_k\to-\infty$.

So only on one end of $\Gamma_{n,n/2}$ (where $t\to+\infty$)
we can have $b\to+\infty$.
On the other hand, each $\Gamma_{n,m},\; m<n/2$ contains
at most two graphs of $\lambda_j$. According to
(\ref{eo}),  the total number of these graphs
$\lambda_j$ is $n+1$, 
and the total number of curves $\Gamma_{n,m}$ is $(n+1)/2$ when $n$ is odd,
and $n/2+1$ when $n$ is even. It follows that, when $n$ is odd, 
each $\Gamma_{n,m}$ contains two graphs of $\lambda_j$.
When $n$ is even, each $\Gamma_{n,m}$
except one contains two graphs of $\lambda_j$, while the exceptional component 
$\Gamma_{n,n/2}$ contains one graph of $\lambda_j$.

Thus $b\to+\infty$ as $c\to\pm1$ in the $X_{k,l}$-charts with $k<0$,
which proves the third statement of Theorem 1.
To prove the last statement,
we study zeros of the eigenfunctions as $b\to+\infty$.

The eigenfunction
$Y_\ell$ of (\ref{harm}) corresponding to the eigenvalue $\mu_\ell$ has
exactly $\ell$ zeros on $i\R$ and no other zeros in $\C$.
One of these zeros is real iff $\ell$ is odd.

The trees corresponding to $Y_\ell$ are constructed similarly
to those corresponding to $y$, using the two loop cell decomposition
of the sphere, consisting of $\gamma_\infty$ and the dashed loop
in Fig.~1, left.
\bigskip
\begin{center}
\epsfxsize=1.0in%
\centerline{\epsffile{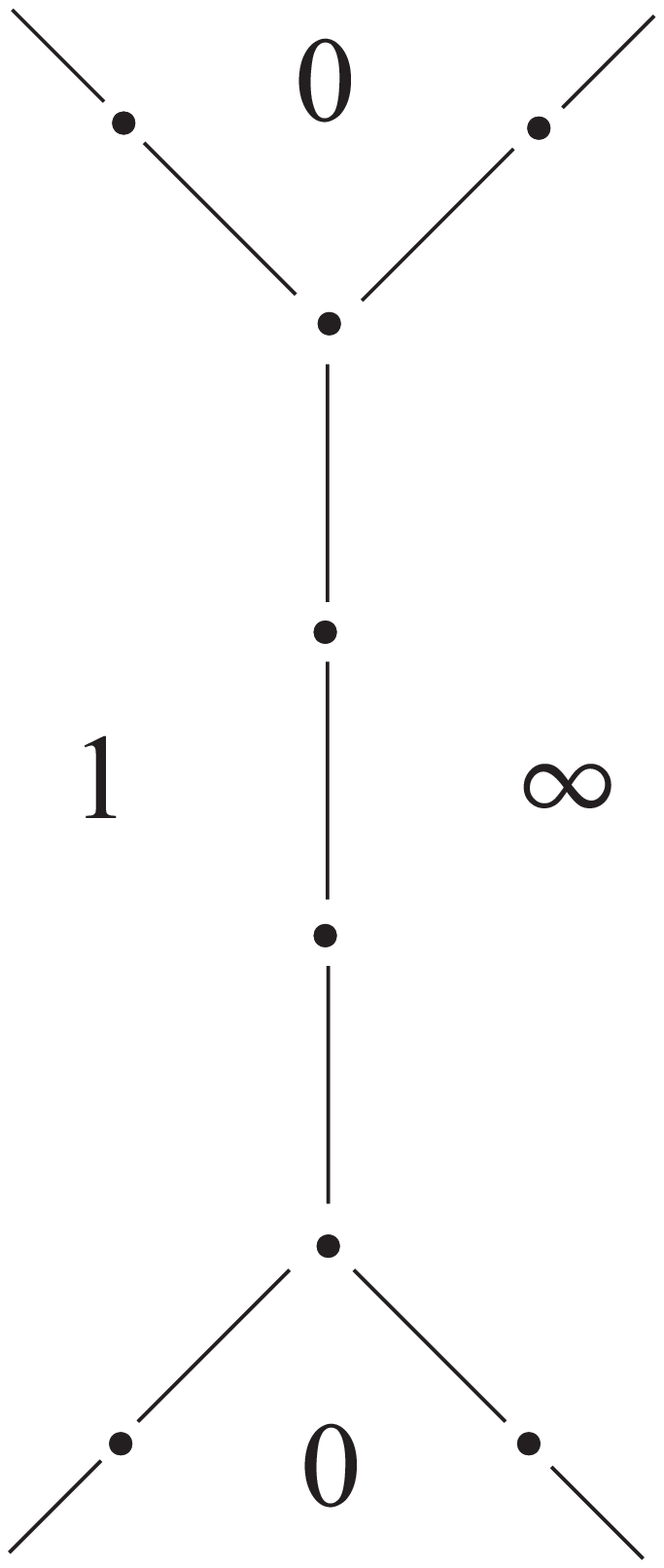}}
\nopagebreak\medskip
\noindent Fig. 11. The tree corresponding to $Y_3$.
\end{center}
\bigskip
For general results on convergence of Nevanlinna functions like
our $f$ we refer to \cite{Volk}.

When $b\to+\infty$, each QES eigenvalue $\lambda(b)$ must tend
to some $\mu_\ell$, and the corresponding QES eigenfunction tends to
$Y_\ell$. Suppose that $\lambda(b)\in\Gamma_{n,m}$ with 
$m<n/2$. Then the tree corresponding to $\lambda(b)$ is
$X_{k,m},\; k<0,\; 2m-k=n$. 
From the arguments in the proofs of Lemmas 1 and 2
(see Figs.~4, 7), degeneration of the cell decomposition $\Psi$
corresponding to such a tree has a connected
component with $2m$ bounded faces when $\beta\to 0$ and with
$2m+1$ bounded faces when $\beta\to\pi$.
This implies that the corresponding eigenfunction can only
converge to $Y_{2m}$ as $\beta\to 0$ and to $Y_{2m+1}$ as
$\beta\to\pi$. 

If $n$ is odd, these curves constitute the whole QES locus.
If $n$ is even, there is one more branch $\lambda(b)$ 
of the QES locus for large positive $b$,
the right end of $\Gamma_{n,n/2}$ corresponding to the 
tree $X_{0,n/2}$. From the proof of Lemma 3 (see Fig.~2, left)
degeneration of the cell decomposition $\Psi$
corresponding to such a tree has a connected
component with $n$ bounded faces when $\beta\to 0$.
This implies that the corresponding eigenfunction can only
converge to $Y_{n}$.

So the ordering of the ends of the curves $\Gamma_{n,m}$ corresponds to
the natural ordering
of the first $n+1$ eigenvalues of the harmonic oscillator.
This completes the proof.

{\em Department of Mathematics

Purdue University

West Lafayette, IN 47907

eremenko@math.purdue.edu

agabriel@math.purdue.edu}

\end{document}